# Investigation of Forming Free Bipolar Resistive Switching Characteristics in Al/Mn$_3$O$_4$/FTO RRAM Device

Vidit Pandey[1], Adiba[1], Tufail Ahmad[1*], Priyanka Nehla[2] and Sandeep Munjal[2,3]

[1] *Department of Physics, Aligarh Muslim University, Aligarh-202002, India.*
[2] *Department of Physics, Indian Institute of Technology, Delhi, Hauz Khas, New Delhi-110016, India.*
[3] *Departments of Physics and Astrophysics, Delhi University, New Delhi, 110007, India*

**Abstract:**
Bipolar resistive switching (BRS) phenomenon has been demonstrated in Mn$_3$O$_4$ using Al (Aluminum)/Mn$_3$O$_4$/FTO (Fluorine doped Tin Oxide) Resistive Random Access Memory (RRAM) device. The fabricated RRAM device shows good retention, non volatile behavior and forming free BRS. The Current-Voltage (I-V) characteristics and the temperature dependence of the resistance (R-T) measurements were used to explore conduction mechanisms and the thermal activation energy (E$_a$). The resistance ratio of high resistance state (HRS) to low resistance state (LRS) is ~10$^2$. The fabricated RRAM device shows different conduction mechanisms in LRS and HRS state such as ohmic conduction and space charge limited conduction (SCLC). The rupture and formation of conducting filaments (CF) of oxygen vacancies take place by changing the polarity of external voltage, which may be responsible for resistive switching characteristics in the fabricated RRAM device. This fabricated RRAM device is suitable for application in future high density non-volatile memory (NVM) RRAM devices.

[*]Corresponding Author. Department of Physics, Aligarh Muslim University, Aligarh-202002, India.

Email addresses: tufailahmadphys@gmail.com, Tel No.: +91-9760452087

# 1. Introduction:

Resistive Random Access Memory (RRAM) devices with fast operation speed, excellent endurance, simple structure, low power consumption, superior scalability, non-destructive readout and good compatibility with conventional complementary metal oxide semiconductors (CMOS) technology [1-3] have been considered promising candidate for future high density non-volatile memory (NVM) devices, Neuromorphic computing [2] and Logic applications [3]. It is worth to notice here that Ferroic tunnel junctions have also shown promising potential for the development of Neuromorphic networks [4]. Most of the RRAM devices are based on simple capacitor like Metal-Insulator/dielectric-Metal (MIM) structure [5] and in this insulator/dielectric layer, a conducting path can be generated through a filament by applied external voltage, which is known as the conducting filament (CF) [6]. Once the CF is formed, it may be ruptured or reformed by next cycle of voltage. The SET and RESET process drive the RRAM devices to low resistance state (LRS) and high resistance state (HRS) respectively [7] and, these LRS and HRS are stable and work as different memory level as ON and OFF respectively [8]. Abrupt changes in stable resistance states under the applied fixed voltage are called resistive switching (RS) and this physical phenomenon is used in non-volatile RRAM devices.

On the basis of switching process, RRAM devices can be roughly categorized into two classes, (I) bipolar resistive switching (BRS) devices and (II) unipolar resistive switching (URS) devices. In BRS devices, the direction of switching depends on the polarity of applied voltage [9], whereas, in URS devices, the switching direction depends on the amplitude of the applied voltage but not on the polarity [11,12]. The BRS devices are generally better than URS devices in terms of device consistency, data retention, endurance, reliability, controllability and storage capacity [2,13] because, there is large fluctuations in switching voltages in URS devices, which leads to overlapping of SET and RESET voltages [12]. Furthermore, according to nature, RRAM devices can also be divided into two types, digital (binary) and analog [13]. In digital RRAM devices, filamentary type switching mechanisms are observed and shows an abrupt changes from HRS to LRS or vice versa, In contrast, the homogenous type of switching occurs and gradual changes exhibit in analog RRAM devices [14]. For many RRAM devices, generally the initial step known as "forming" is important to start the RS operation, which usually requires relatively higher bias voltage to develop the CF between top and bottom electrodes of RRAM devices [15]. Sometimes, there is a risk of permanent damage of RRAM devices by this forming process due to higher value of avalanche current [16]. Thus, it is important to pay attention to fabricate the forming free RRAM devices, which have self compliance feature, more reliability and uniform resistive switching property [17,19,20]. For URS devices, the forming step is necessary, while, in BRS devices, the device structure and materials decide the requirement of the forming processes [11]. Because of these reasons, we concentrated our attention to fabricate forming free BRS RRAM device.

RS phenomenon has been observed in various materials and their composites e.g., 2D materials [17], Organic materials [21,22], Oxides [19], Chalcogenides [24,25], Metal nitrides [21], Polymers and Carbon based materials [20,27] etc. Transition metal oxides (TMOs) like $TiO_2$ [23], NiO [10], ZnO [24], $V_2O_5$ [5], $Al_2O_3$ [11], $NiFe_2O_4$ [15], $CoFe_2O_4$ [25], $HfO_2$ [8] etc. exhibit RS phenomenon and, their RRAM devices perform quiet well in terms of low power consumption, good scalability, device reliability and uniform switching [26]. In TMOs, the RS phenomenon is occurred due to oxygen vacancies and, generation or migration of these vacancies depends upon applied voltage and its polarity [25]. The mobility of oxygen-related defects or vacancies is much higher than that of cations [7]. In most of the oxides based RRAM devices, ohmic conduction, Pool-Frenkal conduction, Schottky emission and space charge limited conduction (SCLC) are commonly observed conduction mechanisms [11]. It is necessary to identify the exact conduction mechanisms and its relationship with different RS properties to increase the performance of the RRAM devices [25]. The RS behaviors are observed in both thin-films [13,31,32] and nanoparticles [17,19,33]. Some groups have reported the study of RS phenomenon in $Mn_3O_4$ thin-film [11,34-36], however, there is no study reported on RS behavior of $Mn_3O_4$ nanoparticles (NPs) yet. $Mn_3O_4$ is widely investigated by researchers due to its interesting applications in scientific and technical fields and works as potential candidate in Nano-electronic devices [28], Sensors [29], Bio-medical research [30], Supercapacitors [31] and many more novel applications. Flexible electronic devices have attracted a great attention in the recent years [32]. The nanoparticles' can easily be used to fabricate flexible devices.

High dielectric constant (k) is one of the basic requirements for any material to exhibit good resistive switching properties. Recently high-k material "$SiO_2$" have been explored for its superior resistive switching memory [33]. $Mn_3O_4$ has high dielectric constant (~5) [34], compared to extensively studied materials for Resistive Switching, such as $SiO_2$ (~3.5) [35], but it has been explored very less, which makes it a good choice to study its resistive switching characteristics.

In the present study, cost effective and eco-friendly sol-gel method [36] [37] was employed for the synthesis of tetragonal $Mn_3O_4$ NPs and confirmed the pure phase of synthesized tetragonal $Mn_3O_4$ NPs by using X-ray diffraction (XRD) analysis. Aluminum (Al) and Fluorine doped Tin Oxide (FTO) were applied at top and bottom electrode respectively in the fabrication of Al/$Mn_3O_4$/FTO RRAM device. The Current-Voltage (I-V) characteristics data was analyzed to find the conduction mechanism in the fabricated RRAM device. The data retention and the program/erase endurance measurements were carried out to check the stability, switching reliability and non volatile characteristics of the fabricated RRAM device. A possible schematic diagram has also been illustrated to explain the behavior of the fabricated RRAM device for different resistance states. **In the present study, $Mn_3O_4$ nanoparticles were used to fabricate the resistive switching device. The process have several benefits, such as (i) the process is cost effective, (ii) it can easily be used for a large scale production of the material, and (iii) it can easily be used to fabricate flexible devices as well, which can be beneficial for commercial applications.**

## 2. Experimental:

The sol-gel route was employed to synthesize the $Mn_3O_4$ NPs. In this method, 0.135 mol of citric acid ($C_6H_8O_7.H_2O$) was mixed in 150 mol of n-propanol and stirred at 500 rounds per minute (rpm) on magnetic hot-plate stirrer for 30 minutes at 40 °C. After this, 0.075 mol of Manganese (II) sulfate ($MnSO_4.H_2O$) was mixed in this solution and stirred with 500 rpm at 40 °C for 30 minutes. The obtained solution was heated at 90 °C in an oven for 8 hours (h) to reduce the extra amount of n-proponol and finally, the obtained powder was dried in vacuum oven at 120 °C for 4 h. The obtained product was then heated at 500 °C for 12 h in ambient environment to get a completely dried nanopowder. In order to fabricate RRAM test device, the thin film of $Mn_3O_4$ NPs was deposited on FTO substrate by spray coating of $Mn_3O_4$ NPs and Al as top electrode was deposited by a thermal evaporator using a shadow mask. The thickness of $Mn_3O_4$ layer in the present study is ~ 800 nm. A glass layer was used to provide the base of the fabricated RRAM test device.

## 3. Result and Discussion:

XRD technique using copper (Cu-Kα ~1.542 Å) as radiation source was adopted to characterize $Mn_3O_4$ NPs, which was recorded between the range of diffraction angle (2θ) from 27º to 67º as shown in Fig. 1(a). The 1st peak (122) is indexed at 28.81º followed by (200), (103), (211), (004), (220), (105), (312), (303), (321), (224) and (400) peaks at 31.04º, 32.09º, 36.04º, 38.21º, 44.38º, 50.76º, 53.82º, 56.18º, 58.59º, 59.98º and 64.48º angles respectively. All the XRD peaks with corresponding angles were perfectly indexed to the tetragonal symmetry of $Mn_3O_4$ with space group I41/amd(#141) [38] and JCPDS Card No.- (01-089-4837) [39], which confirms that the obtained nanopowder is pure phase tetragonal $Mn_3O_4$ NPs. A schematic diagram of the fabricated RRAM device with Al/$Mn_3O_4$/FTO structure has been shown in Fig. 1(b).

[Figure 1 about here.]

In order to investigate the resistive switching phenomenon, the I-V measurement was employed on the fabricated Al/$Mn_3O_4$/FTO RRAM devices with two different thickness of $Mn_3O_4$ layer. As the device with lower $Mn_3O_4$ layer thickness (~ 800 nm) smaller set and reset voltages (inset of Fig. 2(a)), only this device has been studied in the subsequent discussion. The fabricated RRAM device was measured under an external bias voltage that was swept between -5 volt and +5 volt in a cyclic manner without using any compliance current ($I_{cc}$). Fig. 2(a) and Fig. 2(b) show the I-V curves of the device and arrows show the direction of switching between HRS and LRS. Initially, the fabricated RRAM device was in HRS and as increasing the positive bias voltage at top electrode, a clear switching from HRS to LRS was occurred at ~3.7 volt ($V_{SET}$), this switching is referred to SET process. In the similar manner, the switching from LRS to HRS was observed at negative voltage of 4 volt ($V_{RESET}$), when the fabricated device was subjected to negative voltage sweep and this switching is called RESET process. During the SET process, the fabricated RRAM device shows self compliance feature, which reduce the risk of permanent damage of the device due to the higher value of avalanche current. For this device, the RESET and SET powers are ~2.1 mW and ~4.5 mW respectively, which shows the less power consumption capability of the fabricated RRAM device. The ratio from HRS (OFF) to LRS (ON) has been found as ~$10^2$ at read voltage 0.1 volt and the value of nonlinearity factor ($I_{VSET}/I_{0.1Vread}$) is ~ 37 [40]. This indicates that the fabricated RRAM device shows good resistive switching positive bias voltage directions.

[Figure 2 about here.]

To investigate the stability of HRS and LRS states and switching reliability, the data retention and the program/erase endurance measurements were carried out at read voltage, 0.1 volt on the fabricated RRAM device at room temperature (Fig. 3). The values of resistance in HRS and LRS remain constant over a long time period >$10^4$ seconds as shown in Fig. 3(a) and the resistance ratio between HRS (OFF) and LRS (ON) was ~$10^2$, which confirms the stability and non-volatile characteristics of the fabricated Al/$Mn_3O_4$/FTO RRAM device. To examine endurance test, the I-V measurements of the fabricated RRAM device in the cyclic voltage sweep were repeated 200 times at room temperature as shown in Fig. 3(b). Fig. 3(c) shows the histogram presenting "OFF/ON" resistance ratio for 50, 100, 150 and 200 testing cycles, which indicate that, there was no large fluctuation in "OFF/ON" resistance ratio observed with increasing numbers of cycles. The program/erase endurance performance of the fabricated RRAM device reveals a good controllable bipolar resistive switching and stability.

[Figure 3 about here.]

To understand the resistive switching conduction mechanisms of the fabricated RRAM device, the temperature dependence of the resistance (R-T) measurements of the device and the I-V characteristics were studied, when the fabricated RRAM device was in LRS and HRS modes. The results of R-T analysis at 0.1 volt applied bias voltage in temperature range from 300 Kelvin (K) to 400 K are shown in Fig. 4(a). In LRS, the fabricated RRAM device shows metal-like behavior, in contrast, the semiconducting behavior was observed in HRS. The thermal activation energy ($E_a$) was calculated by using Arrhenius equation, $R=R_o exp(E_a/k_B T)$ [41], where, $R_o$, $k_B$ and T are resistance at room temperature, Boltzmann constant and absolute temperature respectively. In HRS, by linear fit of temperature dependence resistance to this equation (Fig. 4(b)), $E_a$ was estimated as ~ 62 meV. By calculating the value of temperature coefficient of resistance (TCR) for LRS, we can identify the nature of conducting filament as we have demonstrated in our previous studies [1,42]. The smaller value of TCR (~ $1\times10^{-3}$ $K^{-1}$ to $2\times10^{-3}$ $K^{-1}$) indicates the presence of conducting filament formed of oxygen vacancies, whereas the relatively larger value of TCR (~ $3.3\times10^{-3}$ K$^{-1}$ to $4\times10^{-3}$ K$^{-1}$) indicates the presence of conducting filament formed of metal atoms [11]. In the present study, in order to find out the value of Temperature coefficient of resistance (α), the resistance vs temperature (R-T) measurements were performed in LRS also (Fig. 4 (c)). The temperature dependent resistance data of LRS was linearly fitted using the equation $R_T = R_0 [1 + α(T − T_0)]$, where $R_T$ and $R_0$ are resistance at temperature T and $T_0$, and α is the temperature coefficient of resistance. The fitted value of TCR (α) is found as ~ $1.93\times 10^{-3}$ K$^{-1}$, which confirms presence of oxygen vacancies' filament in the LRS of current device.

[Figure 4 about here.]

In order to explain further, the conduction mechanisms of resistive switching in the fabricated RRAM device, the I-V characteristic's curve was analyzed in the voltage sequence of 0V → +5V → 0V → -5V → 0V. In the LRS region, a linear I-V plot between ln (I) vs. ln (v), is observed as shown in Fig. 5(a), which confirms the ohmic conduction model due to the formation of well maintain conductive filament during SET process. On the other hand, in HRS region, the asymmetric non-linear I-V characteristics indicate that the conduction mechanism may be Poole-Frenkel conduction [43], Schottky emission [44] or space charge limited current (SCLC) conduction [45] [46]. These different conduction mechanisms follow specific voltage-current relation, for example, Poole-Frenkel, Schottky emission and SCLC conduction are expressed as ln (I/V) vs. $V^{1/2}$, ln (I) vs. $V^{1/2}$ and I vs. $V^2$ respectively [47]. For present fabricated RRAM device, the I-V characteristics data in HRS region can be divided into two regions as low voltage region and high voltage region. The ohmic conduction mechanism's curve can be fitted with straight line in low voltage region (Fig. 5(b)), while, SCLC mechanism's curve can be fitted with straight line in high voltage region as shown in Fig 5(c). In the solid materials, the SCLC is obtained by the injection of electrons from highly carrier injective electrodes at an ohmic contact [46]. Domination of two different type of conduction mechanisms at different bias voltage in HRS region has been observed earlier also in different RRAM devices [25] [48].

In oxide based Resistive Switching (RS) devices some bubbles and cracks may appear on the top electrodes due to the oxygen evolution, on the application of external bias. We did not observe any such deformation of the RS cells in the present device. In our previous studies on $CoFe_2O_4$ (CFO) based RS devices, we have demonstrated that the occurrence of these bubbles/cracks depends upon the choice of top electrodes. For Au/CFO/FTO RS device, oxygen built up near the Au/CFO interface and caused the damage of the device by rupturing the top electrode, whereas no oxygen bubbles and cracks were observed at the top electrodes for Al/CFO/FTO RS device [25]. Shen at al. have also observed similar bubbles/cracks for Pt/BaSrTiO$_3$/SrRuO$_3$ RS device but not for W/BaSrTiO$_3$/SrRuO$_3$ RS device, and suggested the role of top active electrode as a reservoir of oxygen ions in the case of W/BaSrTiO$_3$/SrRuO$_3$ [49]. So, as per our experience and the studies of other researchers, by using a chemically active metal as top electrode (such as Al, W etc.) one can reduce the possibility of such damage of the RS device. As in our case, the top electrode is Al and on applying the positive bias to this, the incoming oxygen anions may oxidize it partially, which serves as an oxygen reservoir and helps in improving the RS characteristics.

[Figure 5 about here.]

A possible schematic representation of the different resistance states in the fabricated Al/Mn$_3$O$_4$/FTO RRAM device is given in Fig. 6. When the external positive voltage was applied to the top electrode (Al), the SET process has been started. In this SET process, the oxygen anions (O$^{2-}$) left their sites and started moving towards top electrode, and the oxygen vacancies were generated on their sites. When these oxygen vacancies were gathered, they aligned and formed a conducting filament between top and bottom electrodes. Thus, the fabricated RRAM device was switched from HRS to LRS. LRS has more oxygen vacancies as compared to HRS. On changing the polarity of applied bias voltage, the oxygen anions were moved back to Mn$_3$O$_4$ NPs layer, which occupied the oxygen vacancies partially below top electrode as shown in Fig. 6 and this process is known as RESET process. In this way, the fabricated RRAM device switched back from LRS to HRS. In HRS, the conducting filament was ruptured as some oxygen vacancies were refilled by oxygen anions and made them to regular oxygen sites. In Fig. 6, the arrows show the cycle of resistive switching as SET and RESET process between HRS and LRS in the present fabricated RRAM device.

[Figure 6 about here.]

**4. Conclusion:**

In summary, bipolar resistive switching phenomenon based Al/Mn$_3$O$_4$/FTO RRAM device with OFF to ON resistance ratio ~$10^2$ was fabricated successfully. XRD analysis confirmed the pure phase of synthesized tetragonal Mn$_3$O$_4$ NPs. The fabricated RRAM device shows good retention, non volatile behavior with good endurance, low power consumption and forming free characteristics. Detailed study of the Current-Voltage (I-V) curves confirmed the ohmic conduction in low resistance state, while ohmic conduction at low voltage region and space charge limited conduction at high voltage region governed the fabricated device in high resistance state. The temperature dependence of the resistance (R-T) measurements of the device gave the thermal activation energy ($E_a$) ~ 62 meV. The fabricated RRAM device shows digital nature due to abrupt changes from HRS to LRS or vice versa. A possible schematic representation to explain the formation and rupture of the conductive filament between top and bottom electrode has also been illustrated. The fabricated Al/Mn$_3$O$_4$/FTO RRAM device and its positive results suggest that Mn$_3$O$_4$ may be promising candidate for the high density non-volatile RRAM devices.

**Acknowledgement:**

Two of the authors, Vidit Pandey and Adiba wish to thank University Grants commission (UGC), Government of India for financial supports in the form of a fellowship.

**Reference:**

[1]   S. Munjal, N. Khare, Multilevel resistive and magnetization switching in Cu/CoFe2O4/Pt device: Coexistence of ionic and metallic conducting filaments, Appl. Phys. Lett. 113 (2018) 243501. doi:10.1063/1.5043078.

[2]   T. Breuer, L. Nielen, B. Roesgen, R. Waser, V. Rana, E. Linn, Realization of Minimum and Maximum Gate Function in Ta2 O5 -based Memristive Devices, Sci. Rep. 6 (2016) 1–9. doi:10.1038/srep23967.

[3]   S. Gao, F. Zeng, M. Wang, G. Wang, C. Song, F. Pan, Implementation of Complete Boolean Logic Functions in Single Complementary Resistive Switch., Sci. Rep. 5 (2015) 15467. doi:10.1038/srep15467.

[4]   R. Guo, W. Lin, X. Yan, T. Venkatesan, J. Chen, Ferroic tunnel junctions and their application in neuromorphic networks, Appl. Phys. Rev. 7 (2020) 11304. doi:10.1063/1.5120565.

[5]   Z. Wan, H. Mohammad, Y. Zhao, C. Yu, R.B. Darling, M.P. Anantram, Engineering of the resistive switching properties in V2O5 thin film by atomic structural transition: Experiment and theory, J. Appl. Phys. 124 (2018) 0–7. doi:10.1063/1.5045826.

[6]   M. Lanza, A review on resistive switching in high-k dielectrics: A nanoscale point of view using conductive atomic force microscope, Materials (Basel). 7 (2014) 2155–2182. doi:10.3390/ma7032155.

[7]   R. Waser, M. Aono, Nanoionics-based resistive switching memories, Nanosci. Technol. A Collect. Rev. from Nat. Journals. (2009) 158–165. doi:10.1142/9789814287005_0016.

[8]   J. Shang, W. Xue, Z. Ji, G. Liu, X. Niu, X. Yi, L. Pan, Q. Zhan, X.-H. Xu, R.-W. Li, Highly flexible resistive switching memory based on amorphous-nanocrystalline hafnium oxide films, Nanoscale. 9 (2017) 7037–7046. doi:10.1039/C6NR08687J.

[9]   R. Waser, R. Dittmann, C. Staikov, K. Szot, Redox-based resistive switching memories nanoionic mechanisms, prospects, and challenges, Adv. Mater. 21 (2009) 2632–2663. doi:10.1002/adma.200900375.

[10]  S. Wu, S. Li, Compliance current dependence of conversion between bipolar, unipolar, and threshold resistance switching in Mn 3 O 4 films, AIP Adv. 5 (2015) 87154. doi:10.1063/1.4929475.

[11] S. Munjal, N. Khare, Advances in resistive switching based memory devices, J. Phys. D. Appl. Phys. 52 (2019) 433002. doi:10.1088/1361-6463/ab2e9e.

[12] H. Abbas, M.R. Park, Y. Abbas, Q. Hu, T.S. Kang, T.S. Yoon, C.J. Kang, Resistive switching characteristics of manganese oxide thin film and nanoparticle assembly hybrid devices, Jpn. J. Appl. Phys. 57 (2018) 06HC03. doi:10.7567/JJAP.57.06HC03.

[13] A.P. Patil, K.A. Nirmal, S.S. Mali, C.K. Hong, T.G. Kim, P.S. Patil, T.D. Dongale, Tuning the analog and digital resistive switching properties of TiO2 by nanocompositing Al-doped ZnO, Mater. Sci. Semicond. Process. 115 (2020) 105110. doi:10.1016/j.mssp.2020.105110.

[14] C. Kumari, I. Varun, S.P. Tiwari, A. Dixit, Interfacial layer assisted, forming free, and reliable bipolar resistive switching in solution processed BiFeO3 thin films, AIP Adv. 10 (2020) 025110 (1-7). doi:10.1063/1.5134972.

[15] S. Munjal, N. Khare, Forming free resistive switching characteristics in Al/NiFe2O4/FTO device, AIP Conf. Proc. 20171 (2020) 20171. doi:10.1063/5.0001806.

[16] S. Munjal, N. Khare, Electroforming free controlled bipolar resistive switching in Al/CoFe 2 O 4 /FTO device with self-compliance effect, Appl. Phys. Lett. 112 (2018) 73502. doi:10.1063/1.4998401.

[17] Q. Zhao, Z. Xie, Y.P. Peng, K. Wang, H. Wang, X. Li, H. Wang, J. Chen, H. Zhang, X. Yan, Current status and prospects of memristors based on novel 2D materials, Mater. Horizons. 7 (2020) 1495–1518. doi:10.1039/c9mh02033k.

[18] R.G. Jadhav, A. Kumar, S. Kumar, S. Maiti, S. Mukherjee, A.K. Das, Benzoselenadiazole-Based Conjugated Molecules: Active Switching Layers with Nanofibrous Morphology for Nonvolatile Organic Resistive Memory Devices, Chempluschem. 85 (2020) 910–920. doi:10.1002/cplu.202000229.

[19] C. Yao, J. Li, S.K. Thatikonda, Y. Ke, N. Qin, D. Bao, Introducing a thin MnO2 layer in Co3O4-based memory to enhance resistive switching and magnetization modulation behaviors, J. Alloys Compd. 823 (2020) 153731. doi:10.1016/j.jallcom.2020.153731.

[20] W. Wang, G.N. Panin, X. Fu, L. Zhang, P. Ilanchezhiyan, V.O. Pelenovich, D. Fu, T.W. Kang, MoS 2 memristor with photoresistive switching, Sci. Rep. 6 (2016) 1–10. doi:10.1038/srep31224.

[21] S. Kim, H. Kim, S. Hwang, M.H. Kim, Y.F. Chang, B.G. Park, Analog Synaptic Behavior

of a Silicon Nitride Memristor, ACS Appl. Mater. Interfaces. 9 (2017) 40420–40427. doi:10.1021/acsami.7b11191.

[22] D. Chaudhary, S. Munjal, N. Khare, V.D.D. Vankar, Bipolar resistive switching and nonvolatile memory effect in poly (3-hexylthiophene) –carbon nanotube composite films, Carbon N. Y. 130 (2018) 553–558. doi:10.1016/j.carbon.2018.01.058.

[23] K.M. Kim, T.H. Park, C.S. Hwang, Dual Conical Conducting Filament Model in Resistance Switching TiO2 Thin Films, Sci. Rep. 5 (2015) 7844. doi:10.1038/srep07844.

[24] N.H. Yan, Y.J. Lin, T.H. Su, H.C. Chang, Optical properties and defects of ZnO nanorods that are modified by treatment with H 2O 2 and used as conductive filaments for poly(methyl methacrylate)-based resistive switching applications, Bull. Mater. Sci. 43 (2020) 1–5. doi:10.1007/s12034-020-2080-1.

[25] S. Munjal, N. Khare, Valence Change Bipolar Resistive Switching Accompanied With Magnetization Switching in CoFe2O4 Thin Film, Sci. Rep. 7 (2017) 12427. doi:10.1038/s41598-017-12579-x.

[26] A.A. Patil, S.S. Khot, R.U. Amate, P.N. Kamble, P.B. Patil, R.N. Bulakhe, I. In, T.D. Dongale, D. kee Kim, Bipolar-resistive switching and memristive properties of solution-processable cobalt oxide nanoparticles, J. Mater. Sci. Mater. Electron. 31 (2020) 9695–9704. doi:10.1007/s10854-020-03515-3.

[27] G.Y. Zhang, D.Y. Lee, I.C. Yao, C.J. Hung, S.Y. Wang, T.Y. Huang, J.W. Wu, T.Y. Tseng, Unipolar resistive switching in ZrO2 thin films, Jpn. J. Appl. Phys. 52 (2013) 41101. doi:10.7567/JJAP.52.041101.

[28] V. Pandey, A. Adiba, S. Munjal, T. Ahmad, Structural and magnetic properties of tetragonal Mn3O4 nanoparticles synthesized by sol-gel method, AIP Conf. Proc. 2220 (2020) 20163. doi:10.1063/5.0001796.

[29] A.M. Gurban, D. Burtan, L. Rotariu, C. Bala, Manganese oxide based screen-printed sensor for xenoestrogens detection, Sensors Actuators, B Chem. 210 (2015) 273–280. doi:10.1016/j.snb.2014.12.111.

[30] V. Pandey, Adiba, S. Munjal, T. Ahmad, Optical properties and spectroscopic investigation of single phase tetragonal Mn3O4 nanoparticles, Mater. Today Proc. 26 (2020) 1181–1183. doi:10.1016/j.matpr.2020.02.238.

[31] D.P.M.D. Shaik, P. Rosaiah, O.M. Hussain, Supercapacitive properties of Mn3O4

nanoparticles synthesized by hydrothermal method, Mater. Today Proc. 3 (2016) 64–73. doi:10.1016/j.matpr.2016.01.122.

[32] X. Yan, Z. Zhou, J. Zhao, Q. Liu, H. Wang, G. Yuan, J. Chen, Flexible memristors as electronic synapses for neuro-inspired computation based on scotch tape-exfoliated mica substrates, Nano Res. 11 (2018) 1183–1192. doi:10.1007/s12274-017-1781-2.

[33] X. Yan, Z. Zhou, B. Ding, J. Zhao, Y. Zhang, Superior resistive switching memory and biological synapse properties based on a simple TiN/SiO2/p-Si tunneling junction structure, J. Mater. Chem. C. 5 (2017) 2259–2267. doi:10.1039/c6tc04261a.

[34] H. Dhaouadi, O. Ghodbane, F. Hosni, F. Touati, Mn3O4 Nanoparticles: Synthesis, Characterization, and Dielectric Properties, ISRN Spectrosc. 2012 (2012) 1–8. doi:10.5402/2012/706398.

[35] J. Robertson, High dielectric constant oxides, Eur. Phys. J. Appl. Phys. 28 (2004) 265–291. doi:10.1051/epjap:2004206.

[36] Adiba, V. Pandey, S. Munjal, T. Ahmad, Structural, morphological and magnetic properties of antiferromagnetic nickel oxide nanoparticles synthesized via sol–gel route, Mater. Today Proc. 26 (2020) 3116–3118. doi:10.1016/j.matpr.2020.02.643.

[37] Adiba, V. Pandey, S. Munjal, T. Ahmad, Structural and optical properties of sol gel synthesized NiO nanoparticles, AIP Conf. Proc. 2270 (2020) 110011. doi:10.1063/5.0020038.

[38] P.R. Garcês Gonçalves, H.A. De Abreu, H.A. Duarte, Stability, Structural, and Electronic Properties of Hausmannite (Mn3O4) Surfaces and Their Interaction with Water, J. Phys. Chem. C. 122 (2018) 20841–20849. doi:10.1021/acs.jpcc.8b06201.

[39] H. Dhaouadi, A. Madani, F. Touati, Synthesis and spectroscopic investigations of Mn3O4 nanoparticles, Mater. Lett. 64 (2010) 2395–2398. doi:10.1016/j.matlet.2010.07.036.

[40] S. Kim, Y. Chang, B. Park, Understanding rectifying and nonlinear bipolar resistive switching characteristics in Ni / SiN x / p-Si memory devices, RCS Adv. 7 (2017) 17882–17888. doi:10.1039/c6ra28477a.

[41] V.R. Rayapati, D. Bürger, N. Du, R. Patra, I. Skorupa, D. Blaschke, H. Stöcker, P. Matthes, S.E. Schulz, H. Schmidt, Electroforming-free resistive switching in yttrium manganite thin films by cationic substitution, J. Appl. Phys. 126 (2019) 74102. doi:10.1063/1.5094748.

[42]  S. Munjal, N. Khare, Compliance current controlled volatile and nonvolatile memory in Ag/CoFe 2 O 4 /Pt resistive switching device, Nanotechnology. 32 (2021) 185204. doi:10.1088/1361-6528/abdd5f.

[43]  S.W. Yeom, S.C. Shin, T.Y. Kim, H.J. Ha, Y.H. Lee, J.W. Shim, B.K. Ju, Transparent resistive switching memory using aluminum oxide on a flexible substrate, Nanotechnology. 27 (2016) 07LT01. doi:10.1088/0957-4484/27/7/07LT01.

[44]  S. Cho, C. Yun, S. Tappertzhofen, A. Kursumovic, S. Lee, P. Lu, Q. Jia, M. Fan, J. Jian, H. Wang, S. Hofmann, J.L. Macmanus-driscoll, Self-assembled oxide films with tailored nanoscale ionic and electronic channels for controlled resistive switching, Nat. Commun. 7 (2016) 12373. doi:10.1038/ncomms12373.

[45]  R.K. Katiyar, Y. Sharma, D.G. Barrionuevo Diestra, P. Misra, S. Kooriyattil, S.P. Pavunny, G. Morell, B.R. Weiner, J.F. Scott, R.S. Katiyar, Unipolar resistive switching in planar Pt/BiFeO 3 /Pt structure, AIP Adv. 5 (2015) 37109. doi:10.1063/1.4914475.

[46]  F.-C. Chiu, A Review on Conduction Mechanisms in Dielectric Films, Adv. Mater. Sci. Eng. 2014 (2014) 1–18. doi:10.1155/2014/578168.

[47]  S. Kundu, M. Clavel, P. Biswas, B. Chen, H.C. Song, P. Kumar, N.N. Halder, M.K. Hudait, P. Banerji, M. Sanghadasa, S. Priya, Lead-free epitaxial ferroelectric material integration on semiconducting (100) Nb-doped SrTiO3 for low-power non-volatile memory and efficient ultraviolet ray detection, Sci. Rep. 5 (2015) 1–14. doi:10.1038/srep12415.

[48]  E. Hernández-Rodríguez, A. Márquez-Herrera, E. Zaleta-Alejandre, M. Meléndez-Lira, W.D. La Cruz, M. Zapata-Torres, Effect of electrode type in the resistive switching behaviour of TiO 2 thin films, J. Phys. D. Appl. Phys. 46 (2013) 45103. doi:10.1088/0022-3727/46/4/045103.

[49]  W. Shen, Investigation of resistive switching in barium strontium titanate thin films for memory applications, 1st ed., Forschungszentrum Jülich, Jülich, 2010. http://juser.fz-juelich.de/record/136188/files/Information_08.pdf (accessed July 15, 2017).

**Figure Captions:**

**Fig.1** (a) X-ray diffraction (XRD) pattern of synthesized single phase tetragonal $Mn_3O_4$ nanoparticles. (b) Schematic diagram of the fabricated RRAM device with $Al/Mn_3O_4/FTO$ structure.

**Fig. 2** (a) Current-Voltage (I-V) curve of the fabricated $Al/Mn_3O_4/FTO$ RRAM device. **Inset shows the variations in SET and RESET voltages with thickness of the $Mn_3O_4$ layer.** (b) The arrows indicate the direction of voltage sweep in the range of -5 volt and +5 volt.

**Fig. 3** (a) Retention properties of the fabricated RRAM device at read voltage 0.1 volt, at room temperature. (b) Endurance performance of the fabricated RRAM device over 200 numbers of cycles at room temperature. (c) Histogram presenting "OFF/ON" resistance ratio for 50, 100, 150 and 200 testing cycles.

**Fig. 4** (a) Resistance vs. Temperature (R-T) curve for High Resistance State at 0.1 volt applied bias voltage. (b) Linear fit of temperature dependence resistance of the Arrhenius equation in High Resistance State. **(c) Linear fit of temperature vs resistance data for the LRS.**

**Fig. 5** Current-Voltage (I-V) curves are shown, (a) Ohmic conduction mechanism in low resistance state (b) Ohmic conduction mechanism at low voltage region in high resistance state and (c) Space charge limited conduction mechanism at high voltage region in high resistance state.

**Fig. 6** Schematic representation of the fabricated $Al/Mn_3O_4/FTO$ RRAM device in low resistance state with positive external bias at top electrode and high resistance state with negative external bias at top electrode. Arrows show the SET and RESET process and small circles represent the oxygen vacancies.

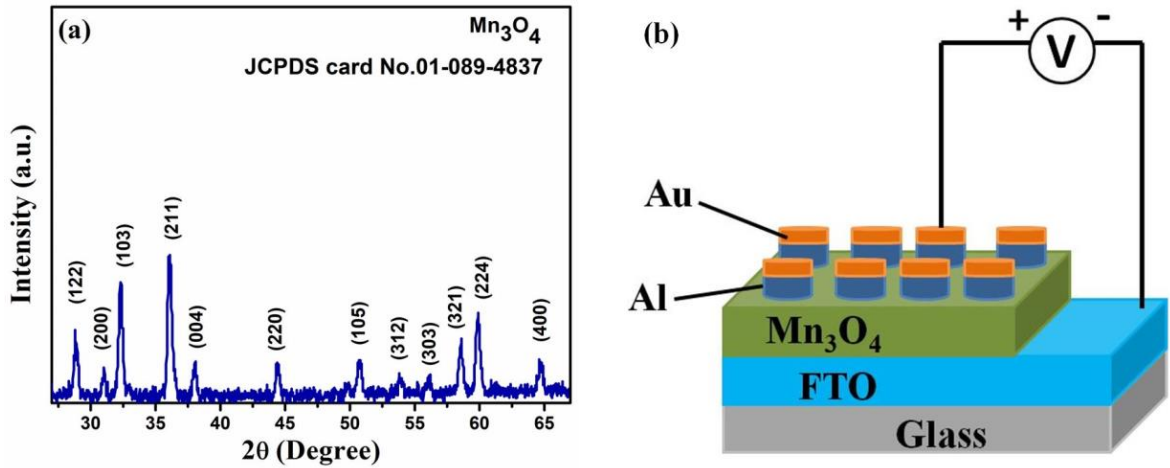

**Fig.1** (a) X-ray diffraction (XRD) pattern of synthesized single phase tetragonal $Mn_3O_4$ nanoparticles. (b) Schematic diagram of the fabricated RRAM device with Al/$Mn_3O_4$/FTO structure.

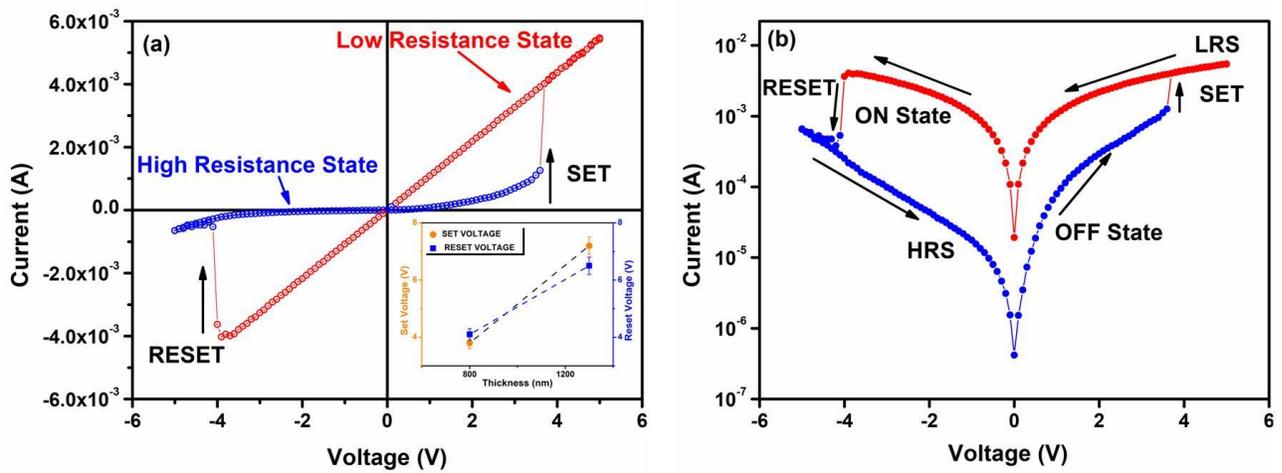

**Fig. 2** (a) Current-Voltage (I-V) curve of the fabricated Al/$Mn_3O_4$/FTO RRAM device. **Inset shows the variations in SET and RESET voltages with thickness of the $Mn_3O_4$ layer.** (b) The arrows indicate the direction of voltage sweep in the range of -5 volt and +5 volt.

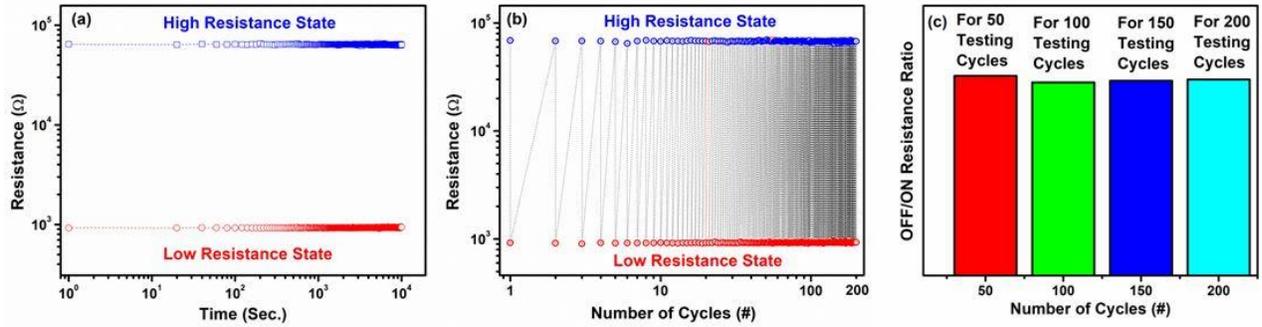

**Fig. 3** (a) Retention properties of the fabricated RRAM device at read voltage 0.1 volt, at room temperature. (b) Endurance performance of the fabricated RRAM device over 200 numbers of cycles at room temperature. (c) Histogram presenting "OFF/ON" resistance ratio for 50, 100, 150 and 200 testing cycles.

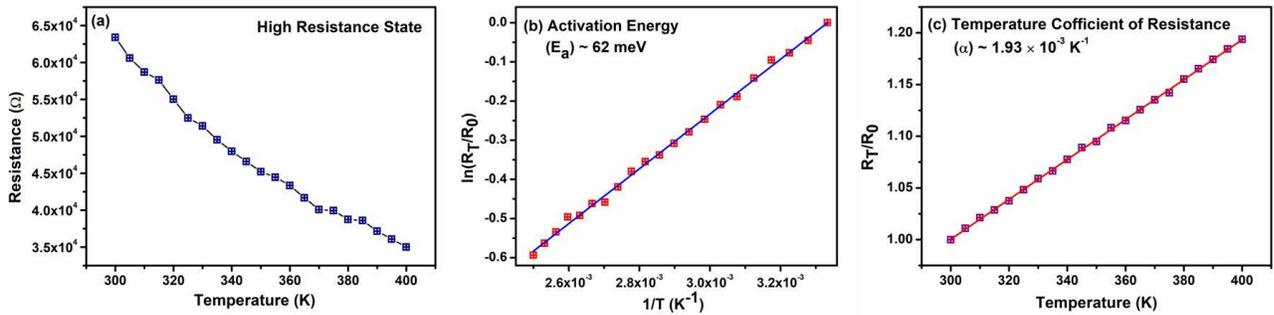

**Fig. 4** (a) Resistance vs. Temperature (R-T) curve for High Resistance State at 0.1 volt applied bias voltage. (b) Linear fit of temperature dependence resistance of the Arrhenius equation in High Resistance State. **(c) Linear fit of temperature vs resistance data for the LRS.**

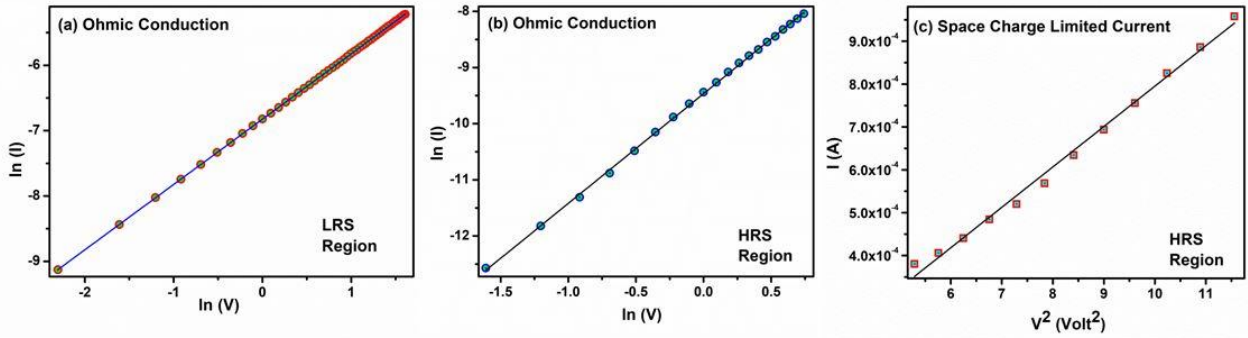

**Fig. 5** Current-Voltage (I-V) curves are shown, (a) Ohmic conduction mechanism in low resistance state (b) Ohmic conduction mechanism at low voltage region in high resistance state and (c) Space charge limited conduction mechanism at high voltage region in high resistance state.

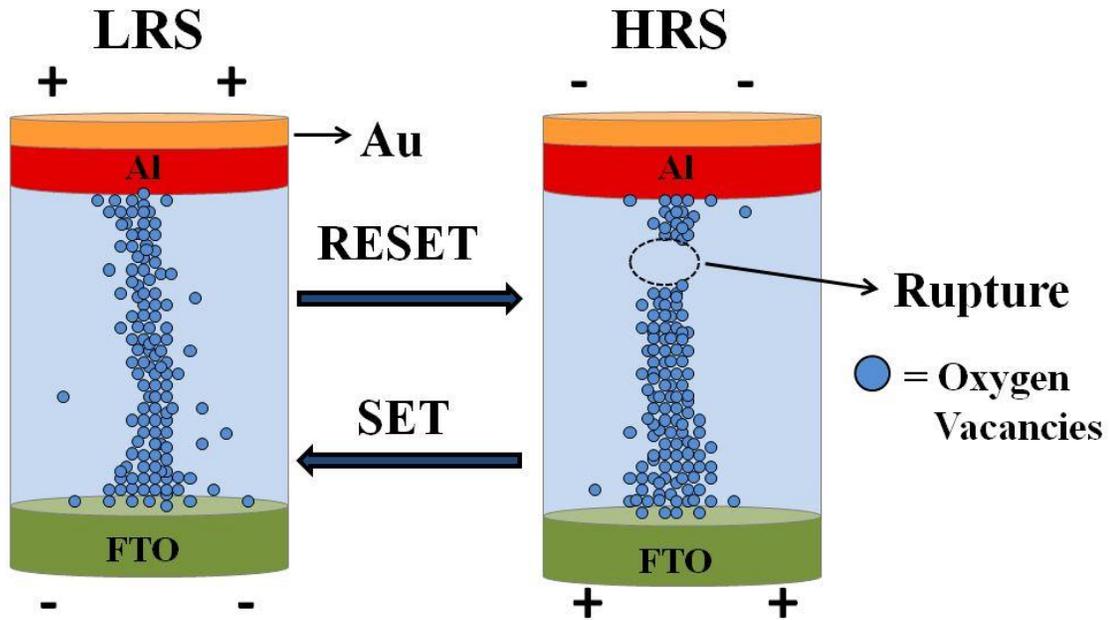

**Fig. 6** Schematic representation of the fabricated Al/Mn$_3$O$_4$/FTO RRAM device in low resistance state with positive external bias at top electrode and high resistance state with negative external bias at top electrode. Arrows show the SET and RESET process and small circles represent the oxygen vacancies.